\documentclass[copyright,creativecommons]{eptcs}
\providecommand{\event}{ACL2 2013} 
\usepackage{breakurl}             

\usepackage{acl2}
\usepackage{xspace}
\usepackage{graphicx}

\usepackage[english]{babel}
\usepackage[utf8]{inputenc}
\usepackage{times}
\usepackage[T1]{fontenc}
\usepackage{lipsum}
\usepackage{ragged2e}
\usepackage{alltt}
\usepackage{rotating}
\usepackage{lscape}
\usepackage{setspace}
\usepackage{array}
\usepackage{verbatim}
\usepackage{fancyvrb}
\usepackage{upquote}
\usepackage{color}

\begin{document}
%

\title{Enhancements to ACL2 in Versions 5.0, 6.0, and 6.1}

\author{
    Matt Kaufmann\\
    \institute{Dept. of Computer Science,\\
               University of Texas at Austin}
    \email{kaufmann@cs.utexas.edu}
    \and
    J Strother Moore\\
    \institute{Dept. of Computer Science,\\
               University of Texas at Austin}
    \email{moore@cs.utexas.edu}
    }

\def\titlerunning{Enhancements to ACL2}
\def\authorrunning{Matt Kaufmann and J Strother Moore}

\maketitle

\begin{abstract}

We report on highlights of the ACL2 enhancements introduced in ACL2
releases since the 2011 ACL2 Workshop.  Although many enhancements are
critical for soundness or robustness, we focus in this paper on those
improvements that could benefit users who are aware of them, but that
might not be discovered in everyday practice.

\end{abstract}




\section{Introduction}

This paper discusses ACL2 enhancements introduced in releases made
since the ACL2 Workshop in November, 2011: Versions 5.0 (August, 2012), 6.0
(December, 2012), and 6.1 (expected February, 2013).  We thus discuss
enhancements made after the release of ACL2 Version 4.3 in July, 2011.

The release notes~\cite{release-notes-doc} for those three versions
report approximately 200 enhancements, which typically were made in
direct response to user feedback or were important to soundness or
robustness of the system.  Our goal in this paper is not simply to
rehash the release notes; rather, it is to highlight important
improvements that ACL2 users are not likely to discover by the routine
use of ACL2.  We do not discuss lower-level improvements to the system
that are reported in comments in source file \texttt{ld.lisp} for the
release notes (e.g., \texttt{(deflabel note-5-0 ...)}).  Those who
dive into the ACL2 sources may wish to peruse these; for example, they
will notice that starting in ACL2 6.0, \texttt{defrec} defines a
recognizer predicate.

Because of the maturity of ACL2, many of the improvements pertain to
aspects of ACL2 that may be unfamiliar to novice users.  Our hope,
however, is that this paper will have value to those users as well, by
suggesting new ideas about what can be done with ACL2.

As in a preceding paper of a similar nature in the previous ACL2
workshop ~\cite{DBLP:journals/corr/abs-1110-4673}, we write ``see :DOC'' to
highlight documentation topics.  For example, see :DOC
\href{http://www.cs.utexas.edu/users/moore/acl2/current/RELEASE-NOTES.html}{release-notes}
and its subtopics (e.g., see :DOC
\href{http://www.cs.utexas.edu/users/moore/acl2/current/NOTE-6-0.html}{note-6-0}
for changes introduced in ACL2 Version 6.0).  Documentation topics are
also referenced implicitly using underlining; for example, the topic
\href{http://www.cs.utexas.edu/users/moore/acl2/current/ADVANCED-FEATURES.html}{\underline{advanced-features}}
provides a handy summary of advanced features of ACL2 in one place.
Each documentation topic reference (of either type) is a hyperlink in
the online version of this paper.

Unlike the preceding paper mentioned above, we choose here to organize
the paper in the way that we have organized the release notes for
several years, as follows.
\begin{itemize}

\item Changes to existing features
\item New features
\item Heuristic improvements
\item Bug fixes
\item Changes at the system level

\end{itemize}

\noindent In each of the five sections corresponding to these topics,
we present a few topics in some detail, but in many cases we simply
note an improvement and point to relevant documentation.

There are typically two other categories: Emacs support and
Experimental/alternate versions.  The former has changed little in the
last few release notes.  As for the latter, there have been
significant improvements to
\href{http://www.cs.utexas.edu/users/moore/acl2/current/HONS-AND-MEMOIZATION.html}{ACL2(h)},
\href{http://www.cs.utexas.edu/users/moore/acl2/current/PARALLELISM.html}{ACL2(p)},
and
\href{http://www.cs.utexas.edu/users/moore/acl2/current/REAL.html}{ACL2(r)};
but for these we live within our space limitations, referring readers
to the release notes.

One outlier, not included in the categories above, is a series of
changes related to licensing and distribution.  For Version 5.0,
changes were made to satisfy University of Texas policies: the license
changed from GPL ``Version 2 or later'' to GPL Version 2, and the
community books~\cite{acl2-books-svn} --- basically, what has been
called the regression suite --- were moved away from the University of
Texas, and are hosted by Google Code.  For Version 6.0 we changed the
license to a BSD-style license, in order to make it easier for
industry groups to take advantage of ACL2.

\subsection*{Acknowledgements}

We thank members of the ACL2 community whose feedback has led us to
continue making improvements to ACL2, including the following, each
mentioned for one or more specific items in the release notes for
Version 5.0, 6.0, or 6.1: Harsh Raju Chamarthi, Jared Davis, Ruben
Gamboa, Shilpi Goel, Dave Greve, David Hardin, Marijn Heule, Warren
Hunt, Anthony Knape, Robert Krug, Camm Maguire, Pete Manolios,
Francisco J. Martin Mateos, David Rager, Jose Luis Ruiz-Reina, Anna
Slobodova, Eric Smith, Rob Sumners, Sol Swords, Sarah Weissman, and
Nathan Wetzler.  We expressly thank Warren Hunt for his continued
support of the use of ACL2, in particular in projects at the
University of Texas.  Finally, we thank the reviewers for helpful
comments, one of which led us to improve :DOC
\href{http://www.cs.utexas.edu/users/moore/acl2/current/PROVISIONAL-CERTIFICATION.html}{provisional-certification}
for the next release.

This material is based upon work supported by DARPA under Contract
No. N66001-10-2-4087, by ForrestHunt, Inc., and by the National
Science Foundation under Grant Nos. CCF-0945316 and CNS-0910913.

\section{Changes to existing features}

There are over 50 release note items about changes to existing features.
Here we list a few and then present a few others in a bit more detail.

\begin{itemize}

\item Functions \texttt{READ-ACL2-ORACLE},
  \href{http://www.cs.utexas.edu/users/moore/acl2/current/READ-RUN-TIME.html}{\underline{\texttt{READ-RUN-TIME}}},
  \texttt{GET-TIMER}, and \texttt{MAIN-TIMER} are no longer
  untouchable; you can call them in your programs.

\item Macros can take an argument named \texttt{STATE}, or which is
  the name of a
  \href{http://www.cs.utexas.edu/users/moore/acl2/current/STOBJ.html}{stobj}.
  However, these variables are not bound to the ``live objects'' as
  you might expect but are treated just like other macro variables.

\item The macros
  \href{http://www.cs.utexas.edu/users/moore/acl2/current/MEMOIZE.html}{\underline{\texttt{MEMOIZE}}}
  and
  \href{http://www.cs.utexas.edu/users/moore/acl2/current/UNMEMOIZE.html}{\underline{\texttt{UNMEMOIZE}}}
  now cause a warning rather than an error in (regular, non-HONS)
  ACL2.

\item The macro
\href{http://www.cs.utexas.edu/users/moore/acl2/current/DEFUND.html}{\underline{\texttt{DEFUND}}}
may now be used without error with
\texttt{:\href{http://www.cs.utexas.edu/users/moore/acl2/current/DEFUND.html}{\underline{PROGRAM}}}
mode specified in an
\href{http://www.cs.utexas.edu/users/moore/acl2/current/XARGS.html}{\underline{\texttt{XARGS}}}
declaration.

\item The functions
\href{http://www.cs.utexas.edu/users/moore/acl2/current/SYS-CALL.html}{\underline{\texttt{SYS-CALL}}}
and
\href{http://www.cs.utexas.edu/users/moore/acl2/current/SYS-CALL-STATUS.html}{\underline{\texttt{SYS-CALL-STATUS}}}
are now
\href{http://www.cs.utexas.edu/users/moore/acl2/current/GUARD.html}{\underline{guard}}-verified
\texttt{:\href{http://www.cs.utexas.edu/users/moore/acl2/current/LOGIC.html}{\underline{\texttt{LOGIC}}}}
mode functions.

\item The environment variable \texttt{ACL2\_COMPILE\_FLG} provides a
default for
\href{http://www.cs.utexas.edu/users/moore/acl2/current/CERTIFY-BOOK.html}{\underline{\texttt{CERTIFY-BOOK}}};
it was formerly named
\texttt{COMPILE\_FLG}.

\end{itemize}

\noindent {\em Some other changes} \\





%


It has been the case since Version 3.6 (August, 2009) that the
definition of a function symbol can mention that symbol in the guard
and measure.  Now, guards specified in
\href{http://www.cs.utexas.edu/users/moore/acl2/current/ENCAPSULATE.html}{\underline{\texttt{ENCAPSULATE}}}
\href{http://www.cs.utexas.edu/users/moore/acl2/current/SIGNATURE.html}{signatures}
may
similarly refer to the functions being introduced in the same
\texttt{ENCAPSULATE} event.

Some utilities have been improved, so you might want to try them again
even if you gave up on them in the past.  For example, consider
\texttt{:\href{http://www.cs.utexas.edu/users/moore/acl2/current/PL.html}{\underline{\texttt{PL}}}}
applied to a non-symbol.  It didn't work for macro calls, but now it
performs macroexpansion (and other transformations to internal form)
as a first step; and moreover, among the rule classes that it shows is
now the \texttt{:LINEAR} class.  Another utility that has been
improved is
\href{http://www.cs.utexas.edu/users/moore/acl2/current/TOP-LEVEL.html}{\underline{\texttt{TOP-LEVEL}}},
which no longer causes calls of
\href{http://www.cs.utexas.edu/users/moore/acl2/current/LD.html}{\underline{\texttt{LD}}}
to stop.  The ``with-error-trace'' utility,
\href{http://www.cs.utexas.edu/users/moore/acl2/current/WET.html}{\underline{\texttt{WET}}},
has also been improved.  Finally, if you haven't yet tried
\href{http://www.cs.utexas.edu/users/moore/acl2/current/DEFATTACH.html}{\underline{\texttt{DEFATTACH}}},
because your code seemed to run a bit slowly using attachments,
consider trying again, as efficiency has improved for this utility.

The abbreviated proof output offered by {\em gag-mode} is now on by
default.  See :DOC
\href{http://www.cs.utexas.edu/users/moore/acl2/current/SET-GAG-MODE.html}{\texttt{SET-GAG-MODE}}
for a description of gag-mode.  If you want a bit of control over the
printing of induction schemes and guard conjectures in gag-mode, see
the discussion of \texttt{:GAG-MODE} in :DOC
\href{http://www.cs.utexas.edu/users/moore/acl2/current/SET-EVISC-TUPLE.html}{\texttt{SET-EVISC-TUPLE}}.

For a macro \texttt{mac}, you can now add a pair \texttt{(mac . fn)}
to the
\href{http://www.cs.utexas.edu/users/moore/acl2/current/MACRO-ALIASES-TABLE.html}{\underline{\texttt{MACRO-ALIASES-TABLE}}}
even when \texttt{fn} has not been
defined as a function symbol.  This can be useful if you want to
define a set of macros early.  See :DOC
\href{http://www.cs.utexas.edu/users/moore/acl2/current/ADD-MACRO-ALIAS.html}{\texttt{ADD-MACRO-ALIAS}}.


%

When functions such as \texttt{FMT-TO-STRING} (see :DOC
\href{http://www.cs.utexas.edu/users/moore/acl2/current/PRINTING-TO-STRINGS.html}{printing-to-strings})
was introduced in Version 4.3, it printed with a right margin set to
10,000, but now the default right margin settings are used.  Thus, for
example, the string returned as shown below had no newline characters
in Version 4.3.  We can return to the default behavior as shown.

{\footnotesize
\begin{verbatim}
ACL2 !>(fmt-to-string "~x0"
                      (list (cons #\0 (make-list 20))))
(0
 "
(NIL NIL NIL NIL NIL NIL NIL NIL NIL NIL
     NIL NIL NIL NIL NIL NIL NIL NIL NIL NIL)
")
ACL2 !>(fmt-to-string "~x0"
                      (list (cons #\0 (make-list 20)))
                      :fmt-control-alist
                      `((fmt-soft-right-margin . 10000)
                        (fmt-hard-right-margin . 10000)))
(81
 "
(NIL NIL NIL NIL NIL NIL NIL NIL NIL NIL NIL NIL NIL NIL NIL NIL NIL NIL NIL NIL)")
ACL2 !>
\end{verbatim}
}

The extended metafunctions have been reworked, with improved handling
of forcing and also with the option of returning a tag-tree.  Also, a
unifying substitution has been added to metafunction contexts,
accessed with function \texttt{MFC-UNIFY-SUBST}.  See :DOC
\href{http://www.cs.utexas.edu/users/moore/acl2/current/EXTENDED-METAFUNCTIONS.html}{extended-metafunctions}).

Printing of numbers now pays attention to the print radix; see
:DOC
\href{http://www.cs.utexas.edu/users/moore/acl2/current/SET-PRINT-RADIX.html}{\texttt{SET-PRINT-RADIX}}.
For example, before Version 6.0 the final value was printed below as
\texttt{ABCD1234}.  Notice the use of \verb|#u| to allow underscores
in numbers, which is new.

{\footnotesize
\begin{verbatim}
ACL2 !>(set-print-base 16 state)
<state>
ACL2 !>(set-print-radix t state)
<state>
ACL2 !>#uxabcd_1234
#xABCD1234
ACL2 !>
\end{verbatim}
}










\section{New features}

Of the approximately 50 release note items about new features, we list a few
here and then elaborate on a few others below.

\begin{itemize}

\item See :DOC
  \href{http://www.cs.utexas.edu/users/moore/acl2/current/PRINT-SUMMARY-USER.html}{\texttt{PRINT-SUMMARY-USER}}
  for a way to add to what is printed in event summaries.

\item Commands
\texttt{:\href{http://www.cs.utexas.edu/users/moore/acl2/current/PL.html}{\underline{PL}}}
and
\texttt{:\href{http://www.cs.utexas.edu/users/moore/acl2/current/PR.html}{\underline{PR}}}
now have analogues in
the
\href{http://www.cs.utexas.edu/users/moore/acl2/current/PROOF-CHECKER.html}{\underline{proof-checker}}.

\item See :DOC
  \href{http://www.cs.utexas.edu/users/moore/acl2/current/PROVISIONAL-CERTIFICATION.html}{provisional-certification}
  for how to certify books in parallel even when they they are ordered
  linearly by
  \href{http://www.cs.utexas.edu/users/moore/acl2/current/INCLUDE-BOOK.html}{\underline{\texttt{INCLUDE-BOOK}}}.

\item ACL2 now supports multiple instances of a stobj (whether
conventional or abstract), known as {\em congruent stobjs}.  See :DOC
\href{http://www.cs.utexas.edu/users/moore/acl2/current/DEFSTOBJ.html}{\texttt{DEFSTOBJ}}
and see :DOC
\href{http://www.cs.utexas.edu/users/moore/acl2/current/DEFABSSTOBJ.html}{\texttt{DEFABSSTOBJ}}.

\item Access to the host Lisp's disassembler is now provided in the
  ACL2 loop by the
\href{http://www.cs.utexas.edu/users/moore/acl2/current/DISASSEMBLE$.html}{\underline{\texttt{DISASSEMBLE\$}}}
utility.

\item See :DOC
  \href{http://www.cs.utexas.edu/users/moore/acl2/current/DEFTHEORY-STATIC.html}{\texttt{DEFTHEORY-STATIC}}
  for a variant of
  \href{http://www.cs.utexas.edu/users/moore/acl2/current/DEFTHEORY.html}{\underline{\texttt{DEFTHEORY}}}
  that behaves the same when a book containing such an event is
  included, as it does when when the book was certified.

\item See :DOC
  \texttt{:\href{http://www.cs.utexas.edu/users/moore/acl2/current/PSOF.html}{PSOF}}
  for a variant of
  \texttt{:\href{http://www.cs.utexas.edu/users/moore/acl2/current/PSO.html}{\underline{PSO}}}
  that directs proof output hidden by
  \href{http://www.cs.utexas.edu/users/moore/acl2/current/GAG-MODE.html}{gag-mode}
  to a file.  Also see :DOC
  \href{http://www.cs.utexas.edu/users/moore/acl2/current/WOF.html}{\texttt{WOF}}
  for a general utility for directing output to a file.

\item A new macro,
\href{http://www.cs.utexas.edu/users/moore/acl2/current/DEFND.html}{\underline{\texttt{DEFND}}}
is just
\href{http://www.cs.utexas.edu/users/moore/acl2/current/DEFN.html}{\underline{\texttt{DEFN}}}
(i.e.,
\href{http://www.cs.utexas.edu/users/moore/acl2/current/DEFUN.html}{\underline{\texttt{DEFUN}}}
with a
\href{http://www.cs.utexas.edu/users/moore/acl2/current/GUARD.html}{\underline{guard}}
of \texttt{T}) plus a
  \href{http://www.cs.utexas.edu/users/moore/acl2/current/DISABLE.html}{\underline{\texttt{disable}}}
  just as
\href{http://www.cs.utexas.edu/users/moore/acl2/current/DEFUND.html}{\underline{\texttt{DEFUND}}}
is
\href{http://www.cs.utexas.edu/users/moore/acl2/current/DEFUN.html}{\underline{\texttt{DEFUN}}}
plus a \texttt{DISABLE}.

\item New utilities
  \href{http://www.cs.utexas.edu/users/moore/acl2/current/ORACLE-FUNCALL.html}{\underline{\texttt{ORACLE-FUNCALL}}},
  \href{http://www.cs.utexas.edu/users/moore/acl2/current/ORACLE-APPLY.html}{\underline{\texttt{ORACLE-APPLY}}},
  and
  \href{http://www.cs.utexas.edu/users/moore/acl2/current/ORACLE-APPLY-RAW.html}{\underline{\texttt{ORACLE-APPLY-RAW}}},
  provide a sort of higher-order capability, by calling a function
  argument on specified arguments.

\item Both \texttt{INLINE} and \texttt{NOTINLINE} declarations are now
supported for the
\href{http://www.cs.utexas.edu/users/moore/acl2/current/FLET.html}{\underline{\texttt{FLET}}}
utility.

\item See :DOC
  \href{http://www.cs.utexas.edu/users/moore/acl2/current/GC-VERBOSE.html}{\texttt{GC-VERBOSE}}
  for how to control, in some host Lisps, the printing of
  garbage-collection messages.

\item The utility
\href{http://www.cs.utexas.edu/users/moore/acl2/current/ADD-MACRO-FN.html}{\underline{\texttt{ADD-MACRO-FN}}},
which is a replacement for \texttt{ADD-BINOP}, lets you choose whether
macros are to be displayed as flat right-associated calls, for
example, \texttt{(append x y z)} rather than \texttt{(append x (append
  y z))}.

\item The new
  \href{http://www.cs.utexas.edu/users/moore/acl2/current/TIME-TRACKER.html}{\underline{\texttt{TIME-TRACKER}}}
  utility supports annotating your programs to display information
  during a computation about elapsed runtime.

\item The {\em{tau system}} is discussed in Section~\ref{heuristic}.


\end{itemize}

\noindent {\em Some other new features} \\

The utility
\href{http://www.cs.utexas.edu/users/moore/acl2/current/DEFUN-NX.html}{\underline{\texttt{DEFUN-NX}}}
has been improved, for example by avoiding
\href{http://www.cs.utexas.edu/users/moore/acl2/current/STOBJ.html}{\underline{stobj}}
restrictions in the \texttt{:LOGIC} component of an
\href{http://www.cs.utexas.edu/users/moore/acl2/current/MBE.html}{\underline{MBE}}
call.  Here is an example from Jared Davis that motivated this change;
note the call of function \texttt{MY-IDENTITY} on a stobj even though
\texttt{MY-IDENTITY} was not declared to take a stobj argument.
{\footnotesize
\begin{verbatim}
(defstobj foo (fld))
(defun-nx my-identity (x) x)
(defun my-fld (foo)
  (declare (xargs :stobjs foo))
  (mbe :logic (my-identity foo)
       :exec (let ((val (fld foo)))
               (update-fld val foo))))
\end{verbatim}
}
\noindent But there now is another way to violate signatures in non-executable
code: by using the utility,
\href{http://www.cs.utexas.edu/users/moore/acl2/current/NON-EXEC.html}{\underline{NON-EXEC}}.
Note that this time, \texttt{MY-IDENTITY} is defined with
\texttt{DEFN} (which is \texttt{DEFUN} with a guard of \texttt{T}),
not by \texttt{DEFUN-NX}.
{\footnotesize
\begin{verbatim}
(defstobj foo (fld))
(defn my-identity (x) x)
(defun my-fld (foo)
  (declare (xargs :stobjs foo))
  (non-exec (my-identity foo)))
\end{verbatim}
}

There have been many improvements to the
\href{http://www.cs.utexas.edu/users/moore/acl2/current/DOCUMENTATION.html}{\underline{documentation}},
but here we focus on two new topics.  The topic
\href{http://www.cs.utexas.edu/users/moore/acl2/current/ADVANCED-FEATURES.html}{\underline{advanced-features}}
summarizes some cool features of ACL2 that might not all be widely
known, yet may be of interest, especially to experienced users.
Another new topic provides a guide to programming with the ACL2
\href{http://www.cs.utexas.edu/users/moore/acl2/current/STATE.html}{\underline{state}};
see :DOC
\href{http://www.cs.utexas.edu/users/moore/acl2/current/PROGRAMMING-WITH-STATE.html}{programming-with-state}.

A new event,
\href{http://www.cs.utexas.edu/users/moore/acl2/current/DEFABSSTOBJ.html}{\underline{\texttt{DEFABSSTOBJ}}},
provides an interface to conventional single-threaded objects known as
{\em abstract stobjs}~\cite{defabsstobj-paper}.  These can provide
advantages over conventional stobjs in several arenas: execution
speed, proof efficiency, use of symbolic simulation, and modularity of
proof development.

ACL2 now provides a way to direct the host Lisp compiler to inline
calls of a given function.  See :DOC
\href{http://www.cs.utexas.edu/users/moore/acl2/current/DEFUN-INLINE.html}{\texttt{DEFUN-INLINE}}.
We expect that you can generally use this utility just as you would
use
\href{http://www.cs.utexas.edu/users/moore/acl2/current/DEFUN.html}{\underline{\texttt{DEFUN}}}
to define a function.  However, we say a bit more, in part to motivate
our design of this utility.  Fundamentally, \texttt{DEFUN-INLINE} is
simply a macro, as we illustrate by expanding a call of this macro.
{\footnotesize
\begin{verbatim}
ACL2 !>:trans1 (defun-inline f (x)
                 (declare (xargs :guard (consp x)))
                 (integerp (car x)))
 (PROGN (DEFMACRO F (X) (LIST 'F$INLINE X))
        (ADD-MACRO-FN F F$INLINE)
        (DEFUN F$INLINE (X)
               (DECLARE (XARGS :GUARD (CONSP X)))
               (INTEGERP (CAR X))))
ACL2 !>
\end{verbatim}
}
\noindent Notice that \texttt{F} is defined to be a macro whose calls expands to a
corresponding calls of a function, \texttt{F\$INLINE}.  The
invocation of
\href{http://www.cs.utexas.edu/users/moore/acl2/current/ADD-MACRO-FN.html}{\underline{\texttt{ADD-MACRO-FN}}}
arranges that theory functions understand \texttt{F} to mean
\texttt{F\$INLINE} and that proof output will display calls of
\texttt{F\$INLINE} as corresponding calls of \texttt{F}.  But why
didn't we simply support the Common Lisp form \texttt{(declaim (inline
  f))}?  The reason is the support that ACL2 provides for undoing.
Imagine that you want \texttt{F} to be inline and then you change your
mind --- or maybe \texttt{F} is defined in a book that you include
locally.  How can we arrange for Common Lisp to undo the directive to
inline calls of \texttt{F}?  Sadly, the Common Lisp
language~\cite{Steele:1990:CLL:95411} does not provide for a way to do that.  The best
we can do is to direct \texttt{F} to be \texttt{notinline} --- but
that could defeat the host Lisp's appropriate inlining of some
subsequent definition of \texttt{F}.  Our solution is {\em always} to
direct inlining for functions whose name ends in the string
\texttt{"\$INLINE"}, and to provide the illusion that we are defining
a function \texttt{F} rather than \texttt{F\$INLINE}, using
\texttt{ADD-MACRO-FN} as discussed above.  Note that analogous
considerations hold for utility
\href{http://www.cs.utexas.edu/users/moore/acl2/current/DEFUN-INLINE.html}{\underline{\texttt{DEFUN-NOTINLINE}}}.





We invite the ACL2 community to help us to convert ACL2 system
functions from
\texttt{:\href{http://www.cs.utexas.edu/users/moore/acl2/current/PROGRAM.html}{\underline{PROGRAM}}}
mode to
\href{http://www.cs.utexas.edu/users/moore/acl2/current/GUARD.html}{\underline{guard}}-verified
\texttt{:\href{http://www.cs.utexas.edu/users/moore/acl2/current/LOGIC.html}{\underline{LOGIC}}}
mode.  This mechanism is described in some detail
in an online document~\cite{how-to-make-patches}.
Here, in brief, are the steps to follow; we would be happy to provide
more details leading to improvement of the online document.

\begin{enumerate}

\item Install a local copy of ACL2, and build it using \texttt{make}.

\item Develop a book that includes
\href{http://www.cs.utexas.edu/users/moore/acl2/current/VERIFY-TERMINATION.html}{\underline{\texttt{VERIFY-TERMINATION}}}
and
\href{http://www.cs.utexas.edu/users/moore/acl2/current/VERIFY-GUARDS.html}{\underline{\texttt{VERIFY-GUARDS}}}
forms for one or more system functions.  For simplicity we assume here
that there is a single such function, \texttt{FN}.

\item When necessary, modify ACL2 definitions in your copy, for
  example by replacing some calls of \texttt{NULL} by corresponding
  calls of \texttt{ENDP} or by adding or modifying guard
  declarations.  Rebuild your local copy of ACL2 using \texttt{make}.

\item Email us your ACL2 changes and your book, and we will do what is
  necessary in order to incorporate your book into the ACL2 community
  books~\cite{acl2-books-svn} and your changes into the ACL2 sources.

\item Henceforth, the default build of ACL2 will accordingly mark
  \texttt{FN} as a guard-verified \texttt{:LOGIC} mode function.

\end{enumerate}
\noindent We, the ACL2 developers, will check each release that such
proofs still go through, using a build that leaves \texttt{FN} in
\texttt{:PROGRAM} mode.


If you have written
\texttt{:\href{http://www.cs.utexas.edu/users/moore/acl2/current/META.html}{\underline{META}}}
rules or
\texttt{:\href{http://www.cs.utexas.edu/users/moore/acl2/current/CLAUSE-PROCESSOR.html}{\underline{CLAUSE-PROCESSOR}}}
rules, you may have been frustrated that your meta functions and
clause processor functions could not assume the correctness of prover
computations, for example as performed using \texttt{MFC-TS} (see :DOC
\href{http://www.cs.utexas.edu/users/moore/acl2/current/EXTENDED-METAFUNCTIONS.html}{extended-metafunctions}).
A new mechanism, designed with Sol Swords, now provides such a
capability; see :DOC
\href{http://www.cs.utexas.edu/users/moore/acl2/current/META-EXTRACT.html}{meta-extract}).
The community book
\href{https://acl2-books.googlecode.com/svn/trunk/clause-processors/meta-extract-simple-test.lisp}{\texttt{clause-processors/meta-extract-simple-test.lisp}} provides illustrative examples.






ACL2 rule names, or
\href{http://www.cs.utexas.edu/users/moore/acl2/current/RUNE.html}{\underline{rune}s},
form the basis of ACL2
\href{http://www.cs.utexas.edu/users/moore/acl2/current/THEORIES.html}{\underline{theories}}.
But runes do not take into account
\href{http://www.cs.utexas.edu/users/moore/acl2/current/MACRO-ALIASES-TABLE.html}{\underline{macro
    aliases}}
for function symbols.  For example, \texttt{(:definition
  binary-append)} is a rune, and you can use \texttt{append} in a
theory expression to abbreviate the set of runes,
\texttt{\{(:definition binary-append), (:induction binary-append)\}};
but you cannot use \texttt{(:definition append)} in a theory
expression.  Now, however, you can use \texttt{(:d append)} in a
theory expression to designate the rune \texttt{(:definition
  binary-append)}.  There are four new such abbreviation mechanisms,
as follows, where \texttt{symb} is a symbol and \texttt{symb'} is the
macro-aliases dereference of \texttt{symb}; e.g.,
\texttt{binary-append} is the macro-aliases dereference of
\texttt{append}, while \texttt{car} is the macro-aliases dereference
of itself.
\begin{itemize}

\item \texttt{(:d symb . r)} designates the rune
  \texttt{(:definition symb' . r)}.

\item \texttt{(:e symb . r)} designates the rune
  \texttt{(:executable-counterpart symb' . r)}.

\item \texttt{(:i symb . r)} designates the rune
  \texttt{(:induction symb' . r)}.

\item \texttt{(:t symb . r)} designates the rune
  \texttt{(:type-prescription symb' . r)}.

\end{itemize}

Take a new look at ACL2 output when you have large case splits, which
in the past could be difficult to debug.  Now, "Splitter Notes" can
help you locate sources of your case splits.  See :DOC
\href{http://www.cs.utexas.edu/users/moore/acl2/current/SPLITTER.html}{splitter}.

\section{Heuristic improvements}\label{heuristic}

As ACL2 is a heuristic theorem prover, it orchestrates many techniques
to support effective automation of reasoning.  The large regression
suite, contributed by many users over about 20 years, has helped to
tune the prover heuristics so that they often need relatively little
of our attention.  However, we have made improvements since Version
5.0 that include avoidance of some rewriting loops, two
strengthenings of type-set reasoning, and tweaks to the heuristics for
automatically expanding recursive function calls during proofs by
induction.  




ACL2 now expands away calls of so-called {\em guard-holders} before
storing induction schemes.  These include
\href{http://www.cs.utexas.edu/users/moore/acl2/current/THE.html}{\underline{\texttt{THE}}}
as well as
all calls of
\href{http://www.cs.utexas.edu/users/moore/acl2/current/RETURN-LAST.html}{\underline{\texttt{RETURN-LAST}}}.
The latter include
\href{http://www.cs.utexas.edu/users/moore/acl2/current/MBE.html}{\underline{\texttt{MBE}}},
\href{http://www.cs.utexas.edu/users/moore/acl2/current/PROG2$.html}{\underline{\texttt{PROG2\$}}},
and
\href{http://www.cs.utexas.edu/users/moore/acl2/current/EQUALITY-VARIANTS.html}{\underline{equality-variants}} ---
for example, a call of
\texttt{MEMBER} expands to the corresponding call of
\texttt{MEMBER-EQUAL}.  Such expansion also occurs before storing
constraints generated by
\href{http://www.cs.utexas.edu/users/moore/acl2/current/ENCAPSULATE.html}{\underline{\texttt{ENCAPSULATE}}}
\href{http://www.cs.utexas.edu/users/moore/acl2/current/EVENTS.html}{\underline{events}}.

We may think of the
\href{http://www.cs.utexas.edu/users/moore/acl2/current/BREAK-REWRITE.html}{\underline{break-rewrite}}
utility as a heuristic, since, when enabled, it chooses debugging
information to display to the user.  This utility had incurred
significant overhead even when disabled, as it is by default.  That
has been fixed, resulting in elimination of more than 10\% of the time
required for an ACL2 regression.




The remainder of this section discusses a feature introduced in
Version 5.0
that contributes to the set of primary prover heuristics: the
\href{http://www.cs.utexas.edu/users/moore/acl2/current/TAU-SYSTEM.html}{\em{tau
      system}}.
This system is a
decision procedure designed to exploit previously
proved theorems about monadic Boolean functions.  The
tau system was extended and improved in Versions 6.0
and 6.1.

The system mines all the axioms, definitions, and
proved rules (of any
\href{http://www.cs.utexas.edu/users/moore/acl2/current/RULE-CLASSES.html}{rule
class}) relating Boolean
function symbols of one argument.  One might think of
these function symbols as recognizing ``soft types''
such as \texttt{integerp}, \texttt{consp},
\texttt{alistp}, \texttt{n32-bit-numberp}, etc.  The
{\em{tau}} of a term is the set of all such recognizers
known to hold of the value of the term.  The tau of a
term is typically computed in a context specifying the
tau of other terms (typically including variables and
subterms).  For example, if an \texttt{IF} has the test
\texttt{(integerp i)}, then when the tau of the true
branch is computed, the variable \texttt{i} is known to
have a tau that contains \texttt{integerp} and all the
recognizers it is known to imply.

For purposes of the tau system, Boolean monadic
functions are tracked, as are equalities and
inequalities with constants.  As of Version 6.1, the
tau system was extended to track intervals.  For
example, the tau for a term might, in addition to
saying that the value of the term is an integer (and
thus also a rational and not a cons), lies in the
interval between 0 and 15 but is not 3 or 7.

Of special importance are {\em{signature rules}} that
allow the tau system to compute the tau of a function
application by computing the tau of the actuals.  Tau
also tracks other forms of rules that relate the known
predicates, and it allows signatures for the various
values returned by multiple-value functions.  The tau
system also provides a way for the user to define,
verify, and install
``\href{http://www.cs.utexas.edu/users/moore/acl2/current/BOUNDERS.html}{bounder}''
functions which can be
used to compute an interval containing a function's
output from the intervals containing its input.

It is possible to prove certain theorems by tau
reasoning alone.  Such formula are often, informally,
thought of as being mere consequences of ``type
checking.''  The tau system is designed to
recognize such formulas rapidly.  It is thought the tau system,
if properly ``programmed'' with rules, will be helpful
in verifying
\href{http://www.cs.utexas.edu/users/moore/acl2/current/GUARD.html}{\underline{guard}}
conjectures.

The tau documentation has grown extensively since
Version 5.0.  We recommend that the interested reader
see :DOC
\href{http://www.cs.utexas.edu/users/moore/acl2/current/INTRODUCTION-TO-THE-TAU-SYSTEM.html}{introduction-to-the-tau-system}.

\section{Bug fixes}

We have continued to improve ACL2 by eliminating more than 50 bugs.
In this section we mention only a few that may have the most effect on
how people use ACL2.

The time reports in event summaries have been much improved.  As far
as we know, they now accurately report runtime (cpu time).  Of course,
you can use the
\href{http://www.cs.utexas.edu/users/moore/acl2/current/TIME$.html}{\underline{\texttt{TIME\$}}}
utility for reports of realtime and
runtime that avoid the accounting done by ACL2.

%


%

The
\href{http://www.cs.utexas.edu/users/moore/acl2/current/FLET.html}{\underline{\texttt{FLET}}}
construct no longer has any requirements for returning stobjs.

\section{Changes at the system level}

In this section we pick a few additions and improvements that are
outside the realm of what one might normally think of as ``ACL2
features''.




The character encoding for reading from files --- and for some host
Lisps also for reading from the terminal --- is now iso-8859-1, also
known as latin-1.  See :DOC
\href{http://www.cs.utexas.edu/users/moore/acl2/current/CHARACTER-ENCODING.html}{character-encoding}.


You can now build the ACL2
\href{http://www.cs.utexas.edu/users/moore/acl2/current/DOCUMENTATION.html}{\underline{documentation}}
locally (using \texttt{make DOC}).  Previously, the graphics had been
omitted when doing so.

If you want to run a parallel regression using `make', you should now
avoid the `\texttt{-j}' option.  Instead, use \texttt{ACL2\_JOBS=$n$}
where $n$ is the maximum number of jobs to run in parallel.  This
change is in support of including the \texttt{centaur/} books in such
regressions.  (Those books had formerly only been certified in
regressions done for ACL2(h); see :DOC
\href{http://www.cs.utexas.edu/users/moore/acl2/current/HONS-AND-MEMOIZATION.html}{hons-and-memoization}.)
Note that you should still use `\texttt{-j}' if you are certifying
books residing in a particular directory, rather than doing a full
regression.


The search button near the top of the
\href{http://www.cs.utexas.edu/users/moore/acl2/}{ACL2 home page} will
lead you to two search utilities: one for the documentation, and one
for the community books.




\section{Conclusion}

We have presented an outline of changes to ACL2 in Versions 5.0, 6.0,
and 6.1.  Our focus has been to describe changes that can affect one's
daily use of ACL2 but might otherwise go unnoticed.  Many more changes
(close to 200 altogether) may be found in the
\href{http://www.cs.utexas.edu/users/moore/acl2/current/RELEASE-NOTES.html}{release
  notes} for these three versions, and many changes at a lower level
are described in comments in the source code for those release notes
(\texttt{(deflabel note-5-0 ...)} etc.).

A critical component in the continued evolution of ACL2 is feedback
from the user community.  We hope that you'll keep that feedback
coming!  Another contribution of the user community is the large body
of Community Books~\cite{acl2-books-svn}, which put demands on the
system and help us to test improvements.  Please keep these coming,
too!

\bibliographystyle{eptcs}

\bibliography{kaufmann-moore}

\end{document}